\documentclass[12pt,showpaces,preprintnumbers,amsmath,amssymb,twocolumn]{iopart}
\usepackage{graphicx}
\usepackage{dcolumn}
\usepackage{bm}
\usepackage{amssymb}
\usepackage{epstopdf}
\begin{document}

\title{Tunable magnetic orders in CePd$_2$As$_{2-x}$P$_x$}

\author{\author{T Shang, Y H Chen, W B Jiang, Y Chen, L Jiao, J L Zhang, Z F Weng, X Lu, H Q Yuan}}
\address{Center for Correlated Matter and Department of Physics, Zhejiang University, Hangzhou, Zhejiang 310027, China}

\ead{hqyuan@zju.edu.cn}

\begin{abstract}
We report the successful synthesis of the polycrystalline compounds CePd$_2$As$_{2-x}$P$_x$ ($0 \leq x \leq 2$) and their physical properties by measuring the transport, magnetic and thermodynamic behaviors as a function of temperature and/or magnetic field. Powder x-ray diffraction (XRD) indicates that CePd$_2$As$_{2-x}$P$_x$ crystallizes in the ThCr$_2$Si$_2$-type tetragonal structure. CePd$_2$As$_2$ exhibits a moderate Sommerfeld coefficient of $\gamma$ $\approx$ 88 mJ/mol-K$^2$, and undergoes an antiferromagnetic (AFM) transition at the N\'{e}el temperature $T_N \approx$ 15 K. Upon substituting As with P, the $T_N$ is nearly unchanged up to $x \simeq 0.6$, while a ferromagnetic (FM) transition develops below $T_N$ for $x \simeq 0.4$. The Curie temperature $T_C$ increases with increasing $x$ and eventually merges with the AFM transition at $x \simeq 0.6$. With further increase of $x$, the system follows typical FM behaviors and its $T_C$ monotonically increases and reaches $T_C \approx 28$ K in CePd$_2$P$_2$. Moreover, a metamagnetic transition is observed in the As-rich samples, but vanishes for $x \geq 0.4$. Such a tunable magnetic ground state may provide an opportunity to explore the possible quantum critical behavior in CePd$_2$As$_{2-x}$P$_x$.
\end{abstract}
\maketitle

\section{Introduction}

The intermetallic compounds $ReT_2X_2$ ($Re$ = rare-earth element; $T$ = transition metal; $X$ = metalloid element) showed remarkably rich physical properties \cite{HF reviews}. In particular, the Ce- and Yb-based compounds have become the prototype for studying heavy fermion (HF) superconductivity and quantum criticality. The ground state of these compounds can be tuned readily by non-thermal control parameters, e.g., magnetic field, hydrostatic pressure or chemical substitution. For example, CePd$_2$Si$_2$ is an antiferromagnet with a N\'{e}el temperature of $T_N \approx$ 10 K, but it becomes superconducting when the applied pressure suppresses the AFM transition \cite{Mathur1998}. On the other hand, CeCu$_2$(Si,Ge)$_2$ exhibits two superconducting domes under pressure \cite{yuan2003}: the low-pressure dome envelops an AFM quantum critical point (QCP), while the high-pressure one is proposed to be associated with a weak valence transition \cite{yuan2003,Holmes2004}. Furthermore, multiple classes of QCPs were also discussed in these series of compounds \cite{si2010heavy}. For example, CeCu$_2$Si$_2$ shows evidence of spin-density-wave (SDW)-type QCP \cite{Arndt2011}, but a scenario of local QCP was proposed for YbRh$_2$Si$_2$ \cite{Paschen2004}. Following the recent observations of a FM QCP in YbNi$_4$(As,P)$_2$ \cite{Steppke2013}, it would be interesting to search for similar correlated FM compounds in the $ReT_2X_2$ families.

The discovery of superconductivity in iron pnictides $A$Fe$_2$As$_2$ ($A$ = Ca, Sr, Ba, and Eu) with a rather high transition temperature $T_{sc}$ further extends the interests in this category of compounds. In these materials, superconductivity appears while the Fe-SDW order is suppressed either by chemical substitution or by external pressure \cite{reviews}. Furthermore, evidences of magnetic QCPs and non-Fermi liquid behaviors were also reported in BaFe$_2$(As,P)$_2$ \cite{Hashimoto2012, Kasahara2010}. These properties remarkably resemble those of HF compounds. However, in iron pnictides such behaviors are mainly determined by the Fe-$3d$ electrons. Further studies on the relevant compounds might provide insights to the evolution from a HF state to high $T_{sc}$ superconductivity, and enable us to compare the physics of $d$- and $f$-electron systems.

The pnictide compounds CePd$_2$As$_2$ and CePd$_2$P$_2$ crystallize in the ThCr$_2$Si$_2$-type structure, which can be viewed as alternating Ce- and PdAs(P)-layers stacking along the $c$ axis [see inset of figure 1(a)] \cite{Quebe1995,jeitschko1983}. No physical properties have been reported for these compounds in literature so far. Here we present a systematic study on the physical properties of the polycrystalline compounds CePd$_2$As$_{2-x}$P$_x$ ($0 \leq x \leq 2$) by measuring the transport, magnetic and thermodynamic properties. A tunable magnetic ground state is observed for CePd$_2$As$_{2-x}$P$_x$, changing from AFM order to FM order when increasing the P-content from $x = 0$ to $x = 2$.

This article is organized as follows: Immediately after a brief introduction, we describe the experimental methods in Sec. 2. The magnetic transitions of CePd$_2$As$_{2-x}$P$_x$ are characterized by measurements of various physical properties in Sec. 3.1-3.4. In Sec. 3.5, we summarize the magnetic phase diagram of CePd$_2$As$_{2-x}$P$_x$ and discuss its possible origins, followed by a short conclusion in Sec. 4.

\section{Experimental Details}

Polycrystalline samples CePd$_2$As$_{2-x}$P$_x$ ($0 \leq x \leq 2$) were synthesized by a solid-state-reaction method in evacuated quartz ampoules. The Ce (99.9$\%$) and As (99.999$\%$) chunks, Pd (99.95$\%$) powders and P (99.999$\%$) lumps were used as raw materials. First, the precursors of CeAs and CeP were prepared in a similar procedure as described elsewhere \cite{tianGd}. The stoichiometric amounts of CeAs, Pd, As and P were installed in Al$_2$O$_3$ crucibles which were sealed in an evacuated quartz. Then, the quartz ampoules were slowly heated to 1200 K and kept at this temperature for 7 days. After that, the samples were thoroughly ground and pressed into pellets, followed by an annealing at 1200 K again. The preparation process was protected in an argon-filled glove box.

The crystal structure of the derived polycrystalline samples was characterized by powder XRD on a PANalytical X'Pert MRD diffractometer with Cu $K \alpha$ radiation and a graphite monochromator. Magnetic measurements were performed in a Quantum-Design magnetic property measurement system (MPMS-5T). The specific heat measurements were carried out in a Quantum-Design physical property measurement system (PPMS-9T). Temperature dependence of the electrical resistivity was measured by a standard four-point method in an Oxford Instruments HELIOX VL system using a Lakeshore ac resistance bridge.

\begin{figure}[tbp]
     \begin{center}
     \includegraphics[width=3.4in,keepaspectratio]{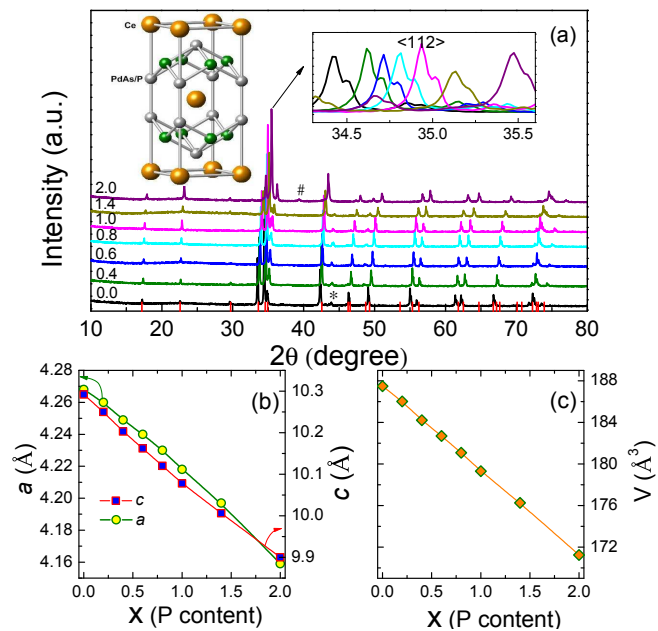}
     \end{center}
     \caption{Powder XRD patterns and lattice parameters for CePd$_2$As$_{2-x}$P$_x$. (a) Room temperature XRD patterns for $x = 0.0, 0.4, 0.6, 0.8, 1.0, 1.4, 2.0$. The insets show the crystal structure (left) and the enlarged $<$\textbf{112}$>$ peaks (right). The lattice parameters (b) and the unit cell volume (c) are plotted as a function of the nominal P-content $x$.}
\end{figure}

\section{Results and Discussion}

\subsection{\label{sec:level2}Crystal structure}

\begin{figure}[bp]
     \begin{center}
     \includegraphics[width=3.4in,keepaspectratio]{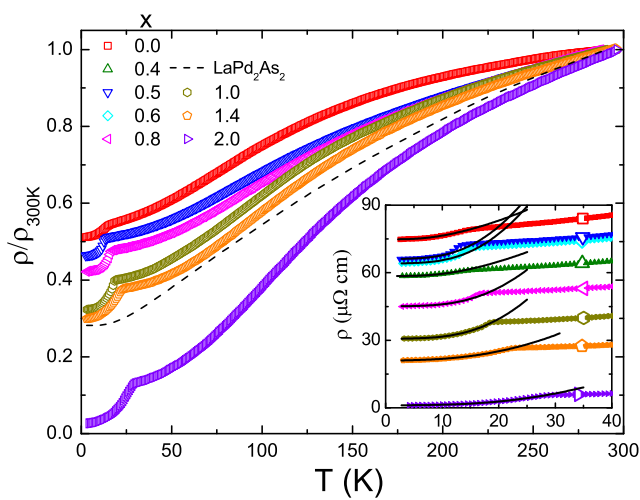}
     \end{center}
     \caption{Temperature dependence of the electrical resistivity for CePd$_2$As$_{2-x}$P$_x$ and LaPd$_2$As$_2$ (dashed line). The inset shows the absolute resistivity $\rho(T)$ in the low temperature region, together with the fits to Eq. (1) ($x < 0.5$) and Eq. (2) ($x \geq 0.5$) (solid lines).}
\end{figure}

Figure 1(a) shows several representative XRD patterns for the polycrystalline CePd$_2$As$_{2-x}$P$_x$ ($0 \leq x \leq 2$), with other samples displaying similar patterns. All the patterns can be well indexed based on the ThCr$_2$Si$_2$-type tetragonal structure with the space group I4/mmm, as shown in the left inset of figure 1(a). The vertical bars on the bottom denote the calculated positions of Bragg diffractions for CePd$_2$As$_2$.  Tiny amount of foreign phases CePd$_3$As$_2$ (marked by $\ast$) and Pd$_5$P$_2$ (marked by $\#$) are detected for $x \leq 1.4$ and $x = 2$, respectively. CePd$_3$As$_2$ shows a simple metallic behavior without any phase transition down to 4.2 K \cite{Gordon1996}. Upon increasing the P-concentration, the $<$\textbf{112}$>$ peak monotonically shifts toward larger $2\theta$ angle [see right inset of figure 1(a)], indicating a continuous contraction of the crystal lattice with increasing $x$. Note that the small satellite peaks are diffractions from Cu $K \alpha2$. The refined lattice parameters decrease with increasing P-content, as shown in figure 1(b). For CePd$_2$As$_2$ and CePd$_2$P$_2$, the lattice parameters of $a = 4.268 \textup{\r{A}}$, $c = 10.292 \textup{\r{A}}$  and  $a = 4.159 \textup{\r{A}}$, $c = 9.901 \textup{\r{A}}$ are in good agreement with previous report \cite{Quebe1995,jeitschko1983}. Figure 1(c) shows a monotonic decrease of the unit cell volume with increasing $x$, which is  reduced by 9.5$\%$ at $x = 2$.

\subsection{Electrical resistivity}

\begin{figure}[bp]
     \begin{center}
     \includegraphics[width=3.4in,keepaspectratio]{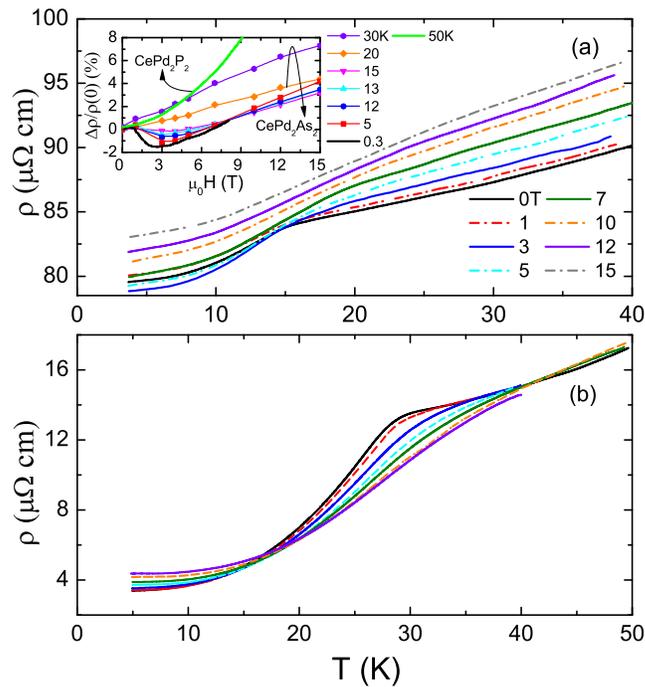}
     \end{center}
     \caption{Temperature dependence of the electrical resistivity $\rho(T)$ at various magnetic fields for (a) CePd$_2$As$_2$ and (b) CePd$_2$P$_2$. Inset of (a) plots the MR, $\Delta\rho/\rho(0)$, of CePd$_2$As$_2$ and CePd$_2$P$_2$ at various temperatures.}
\end{figure}

Figure 2 presents the electrical resistivity $\rho(T)$ of CePd$_2$As$_{2-x}$P$_x$ ($0 \leq x \leq 2$) and its nonmagnetic analog LaPd$_2$As$_2$. All the samples demonstrate metallic behavior down to the lowest temperatures. Upon increasing the P-content $x$, the electrical resistivity is reduced, while the residual resistivity ratio (RRR) is largely enhanced (RRR = 2 for CePd$_2$As$_2$ and RRR = 38 for CePd$_2$P$_2$), demonstrating a better metallicity on the P-side. Such a behavior seems to be caused by the enhanced itinerancy of Ce-4f electrons upon P/As substitution.

At low temperatures, the electrical resistivity $\rho(T)$ of CePd$_2$As$_{2-x}$P$_x$ undergoes a sudden decrease, signaling the onset of a magnetic transition associated with the Ce-moments (see below). The transition temperature is nearly unchanged for $x \leq 0.6$, but increases with further increasing the P-content, reaching $T_C \approx 28$ K at $x = 2$. In the following sections, we will demonstrate that the resistive transition is an AFM-type with $T_N \approx$ 15 K for $x \leq 0.5$, but converts to a FM-type for $x \geq 0.6$. In the magnetically ordered state, the electrical resistivity follows a magnon-gapping behavior. Below the transition temperatures, the electrical resistivity can be well described by the following equations based on the spin wave theory \cite{Baumbach2012,Fontes1999,Yamada1973}:

\begin{equation}
\rho = \rho_0 + A T^2 + B \Delta^2 \sqrt{\frac{T}{\Delta}} [1 + \frac{2}{3}\frac{T}{\Delta} + \frac{2}{15}(\frac{T}{\Delta})^2] e^{-\frac{\Delta}{T}} ~~\textup{(AFM)},
\end{equation}
\begin{equation}
\rho = \rho_0 + A T^2 + B \Delta T (1 + 2\frac{T}{\Delta}) e^{-\frac{\Delta}{T}}  ~~\textup{(FM)},
\end{equation}
where $\rho_0$ is the residual resistivity, and $AT^2$ describes the electron-electron scattering following the Fermi liquid theory. The third term comes from the electron-magnon scattering, where $B$ is a constant and $\Delta$ is the magnon gap. The equations (1) and (2) are for the cases of AFM and FM order, respectively. The solid lines in the inset of figure 2 are fits to Eq. (1) ($x < 0.5$) and Eq. (2) ($x \geq 0.5$), respectively. The derived $\rho_0$, $A$ coefficient and magnon gap $\Delta$ are summarized in the phase diagram (see subsection 3.5).

Figure 3 shows the electrical resistivity of CePd$_2$As$_2$ and CePd$_2$P$_2$ in a magnetic field up to 15 T. For CePd$_2$As$_2$, the resistive transition hardly changes at low fields, but is broadened and shifts to higher temperatures when ramping the magnetic field above 1 T, as a result of the field induced metamagnetic transition. In the inset of figure 3(a), we plot the magnetoresistance (MR), $\Delta\rho/\rho(0)$, of CePd$_2$As$_2$  and CePd$_2$P$_2$ at various temperatures, where $\Delta\rho = \rho(H) - \rho(0)$. The MR at $T = 0.3$ K (CePd$_2$As$_2$) and  $T = 50$ K (CePd$_2$P$_2$) was measured by sweeping the magnetic field and others are derived from the $\rho(T)$ data. For CePd$_2$As$_2$, in the AFM state ($T\leq$ 15 K), the MR shows a local maximum at low fields due to the metamagnetic transition, similar to that observed in EuFe$_2$As$_2$ \cite{Jiang2008}. When the Ce-moments are aligned along the magnetic field, the electrical resistivity increases monotonically with magnetic field, the MR reaching about $3\%$ at 15 T. For CePd$_2$P$_2$, application of a magnetic field broadens the resistive transition and gives rise to a negative MR around $T_C$, as expected for a FM compound. In the paramagnetic (PM) state, the orbital motion of electrons in magnetic field leads to a weak positive MR both in CePd$_2$As$_2$ and CePd$_2$P$_2$.

\subsection{Magnetic properties}

\begin{figure}[tbp]
     \begin{center}
     \includegraphics[width=3.2in,keepaspectratio]{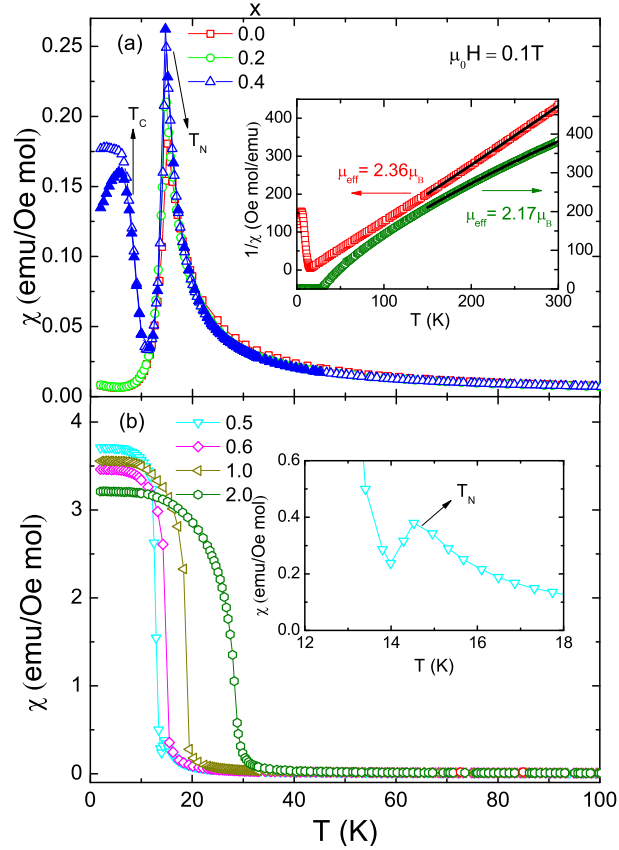}
     \end{center}
     \caption{Temperature dependence of the dc magnetic susceptibility $\chi(T)$ for CePd$_2$As$_{2-x}$P$_x$: (a) $0 \leq x \leq 0.4$; (b) $0.5 \leq x \leq 2$. For $x < 0.4$, no difference was observed between the zero-field-cooling (ZFC) (filled symbols) and field-cooling (FC) (open symbols) data. While for $x \geq 0.4$, the ZFC and FC-curves become separated at low temperatures. The upper inset plots the inverse susceptibility 1/$\chi$ as a function of temperature for $x = 0$ and $2$ . The black lines are fits to the modified Curie-Weiss law. The lower inset expands the magnetic susceptibility $\chi(T)$ around the AFM transition for $x = 0.5$.}
\end{figure}

Figure 4 presents the temperature dependence of the dc magnetic susceptibility $\chi(T)$ for CePd$_2$As$_{2-x}$P$_x$, measured in a field of $\mu_0 H = 0.1$ T. For $x < 0.4$, the magnetic susceptibility $\chi(T)$ exhibits a giant peak around $T_N \approx$ 15 K, which corresponds to the AFM transition. For $x = 0.4$, the magnetic susceptibility $\chi(T)$ shows a steep increase upon cooling below 10 K and it behaves distinctly for the ZFC- and FC-data below 6 K [figure 4(a)], indicating the development of a FM ground state. With further increasing $x$, the magnetic order changes dramatically into a typical FM behavior. As shown in figure 4(b), the FC-susceptibility rises sharply at $T_C$ which increases with $x$ and reaches $T_C \approx$ 28 K in CePd$_2$P$_2$. For $x = 0.5$, an AFM transition is still visible above $T_C$, even though its intensity is greatly weakened [inset of figure 4(b)]. In the paramagnetic state, the magnetic susceptibility can be described by the modified Curie-Weiss law: $\chi(T) = \chi_0 + C/(T - \theta_p)$, with $\chi_0$ being a temperature independent susceptibility including contributions from the core diamagnetism, the van Vleck paramagnetism and the Pauli paramagnetism, $C$ the Curie constant and $\theta_p$ the PM Curie temperature. For example, the inset of figure 4(a) plots the inverse susceptibility $1/\chi(T)$ versus T for CePd$_2$As$_2$ and CePd$_2$P$_2$, respectively, where the black lines are fits to the above Curie-Weiss law. Such a fit allows determining the effective magnetic moment $\mu_{eff}$ from the Curie constant $C$. It is noted that, for CePd$_2$P$_2$, the fitting parameters $\chi_0$ and $C$ may strongly depend on the fitting temperature range; the derived $\mu_{eff}$ decreases but $\chi_0$ increases with lowering the fitting temperature range. This is attributed to the enhanced itinerancy of Ce-$4f$ electrons at low temperatures in CePd$_2$P$_2$, as also evidenced in the electrical resistivity. On the other hand, in CePd$_2$As$_2$ the fitting parameters are much less sensitive to the fitting temperature range and the derived $\chi_0$ is negligible. At sufficiently high temperatures, the derived effective moment $\mu_{eff}$ is comparable to the free-ion moment of Ce (2.54$\mu_B$). For example, we obtained $\mu_{eff}$ = 2.36$\mu_B$/Ce for CePd$_2$As$_2$ and $\mu_{eff}$ = 2.17$\mu_B$/Ce for CePd$_2$P$_2$ while fitting the magnetic susceptibility over a temperature range of $T$ = 150-300 K.
\begin{figure}[htbp]
     \begin{center}
     \includegraphics[width=3.4in,keepaspectratio]{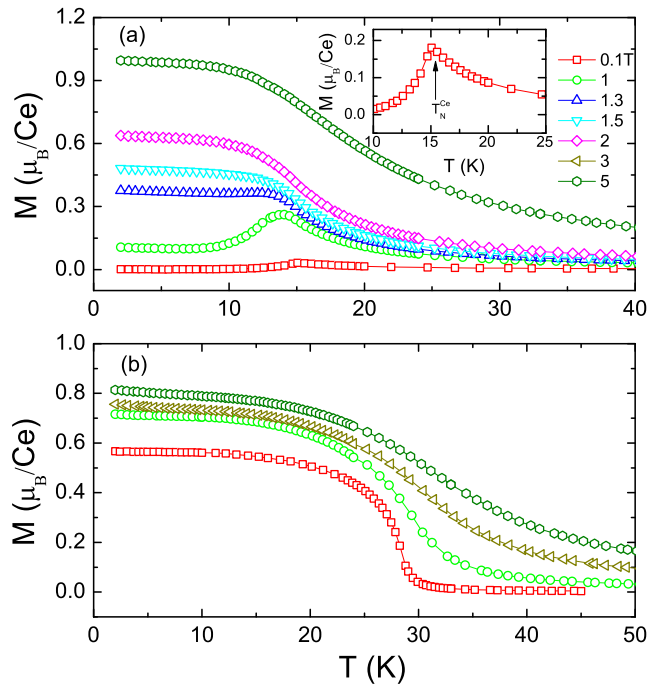}
     \end{center}
     \caption{Temperature dependence of the magnetization $M(T)$ for CePd$_2$As$_2$ (a) and CePd$_2$P$_2$ (b) at various magnetic fields up to 5 T. The inset expands the magnetic transition at $\mu_0 H = 0.1$ T.}
\end{figure}

In order to confirm the field-induced metamagnetic transition in CePd$_2$As$_2$, we plot the temperature dependent magnetization $M(T)$ at various magnetic fields in figure 5. At $\mu_0H =$ 0.1 T, the $M(T)$ curve shows a peak at $T_N \approx$ 15 K [inset of figure 5(a)]. With increasing magnetic field, the magnetic transition in $M(T)$ switches from a peak structure to a step-like behavior, indicating a transition to FM order above 1 T. In contrast, for the FM compound CePd$_2$P$_2$, the applied magnetic field quickly broadens the magnetic transition and shifts it to higher temperatures, being typical of a FM system.

The metamagnetic transition in CePd$_2$As$_{2-x}$P$_x$ ($x < 0.4$) and the emergence of ferromagnetism can be further inferred from the isothermal magnetization $M(H)$. In figure 6, we plot the magnetization $M(H)$ at various temperatures for several doping concentrations $x$. For $x < 0.4$, the magnetization $M(H)$ at temperatures below  $T_N$ initially shows a linear field dependence, followed by a step-like increase at $\mu_0H \approx 0.5$ T and then a saturation above 2 T. Note that the same behavior was also found in CePd$_2$As$_2$ single crystals (not shown). Such a behavior marks a typical metamagnetic transition for CePd$_2$As$_{2-x}$P$_x$ ($x < 0.4$), which is consistent with the results of the electrical resistivity $\rho(T)$ and the magnetization $M(T)$ in the preceding sections. Similar behaviors were also observed in other compounds, e.g., $Re_2$RhIn$_8$, $Re_2$CoGa$_8$, $Re$ = Tb, Dy, Ho \cite{Joshi2008,Cermak2012}. Upon further increasing the P-content ($x \geq 0.4$), the system shows a FM ground state. Correspondingly, the step-like transition in $M(H)$ disappears and spontaneous magnetization develops below $T_C$. The coercive force, determined from the hysteresis loops, increases with $x$. The saturated moment reaches about $1 \mu_B/\textup{Ce}$ at 2 K, which is less than half of that for free Ce-ion ($gJ\mu_B = 2.14\mu_B$) and shows a weak dependence on $x$. This immobile saturated moment also gives additional evidence for the field-induced FM state in samples with low P-contents. Such a strongly reduced moment is likely attributed to the crystal electric field (CEF) effects. Above the magnetic transition temperatures, the magnetization $M(H)$ follows a linear field dependence as expected for the PM state.

\begin{figure*}[htbp]
     \begin{center}
     \includegraphics[width=6.0in,keepaspectratio]{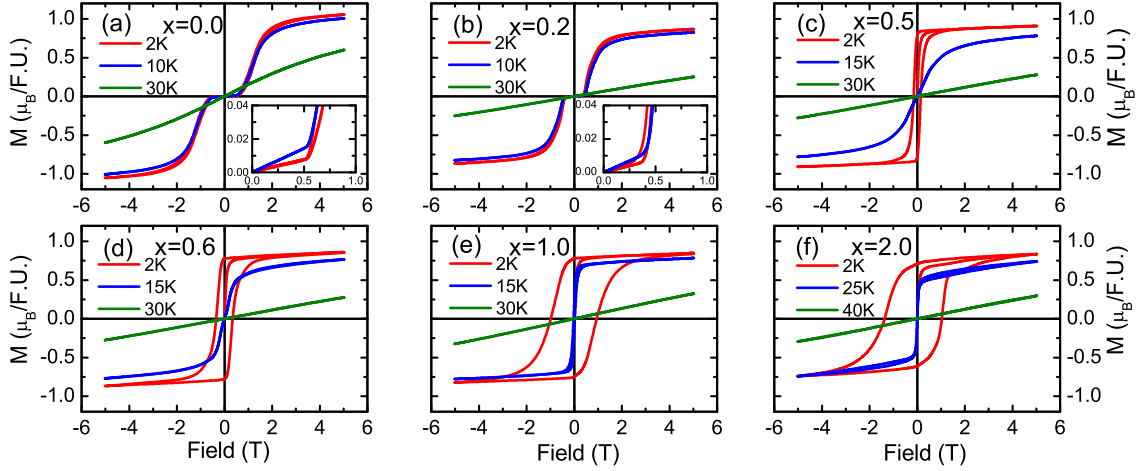}
     \end{center}
     \caption{Field dependence of the magnetization $M(H)$ at several selected temperatures for CePd$_2$As$_{2-x}$P$_x$, $0 \leq x \leq 2$. Insets of (a) and (b) expand the low fields region.}
\end{figure*}

\subsection{Heat capacity}

To further characterize the low-temperature magnetic properties, we also performed the heat capacity measurements for CePd$_2$As$_{2-x}$P$_x$ ($0 \leq x \leq 2$). Figure 7 shows the specific heat divided by temperature, $C/T$, as a function of temperature on a semi-log scale for several representative concentrations. A pronounced $\lambda$-like transition is observed at $T_N \approx$ 15 K for CePd$_2$As$_2$, which is consistent with the AFM transition shown in other measurements. Upon substituting As with P, the magnetic transition remains unchanged up to $x = 0.6$, but shifts to higher temperatures with further doping. A detailed examination shows that there exists a second transition at lower temperatures in the specific heat curves of $x = 0.4$ and $0.5$ [inset of figure 7], corresponding to a subsequent FM transition as evidenced in the magnetic measurements. Note that the specific heat hump around 60 K may result from the CEF effect.

In a magnetically ordered state, the total heat capacity $C$ can be expressed as:
\begin{equation}
C = \gamma T + \beta T^3 + C_{m},
\end{equation}
where $\gamma T$, $\beta T^3$ and $C_{m}$ represent the electronic, lattice and magnetic contributions, respectively. For magnons with a spin gap of $\Delta$, the magnetic contribution $C_m$ is given by \cite{tari2003,continentino2001}

\begin{equation}
C_{m} = aT^{\frac{1}{2}}\Delta^{\frac{7}{2}}[1 + \frac{39}{20}\frac{T}{\Delta} + \frac{51}{32}(\frac{T}{\Delta})^2]e^{-\frac{\Delta}{T}}  ~~\textup{(AFM)},
\end{equation}

\begin{equation}
C_{m} = a(\frac{\Delta^2}{\sqrt{T}}  + 3\Delta \sqrt{T} + 5 \sqrt{T^3}) e^{-\frac{\Delta}{T}}  ~~\textup{(FM)},
\end{equation}
with $a$ being a constant. In Figure 7, the solid lines present fits to Eq. 3, and the derived magnon gaps are shown in the phase diagram [figure 8(a)]. It is noted that the analysis for $x = 0.4$ is complicated by its two subsequent magnetic transitions. Following the above analysis, we also obtained the Sommerfeld coefficient $\gamma$ of CePd$_2$As$_{2-x}$P$_x$, which are summarized in figure 8(b) as a function of P-content $x$. The $\gamma$ value roughly decreases with increasing P-content and exhibits a weak anomaly around $x = 0.5$, being consistent with the resistivity coefficients $A(x)$. For CePd$_2$As$_2$ and CePd$_2$P$_2$, the derived $\gamma$ value are 88 mJ/mol-K$^2$ and 46 mJ/mol-K$^2$, respectively. Such moderate $\gamma$ value suggests a significant hybridization effect in CePd$_2$As$_{2-x}$P$_x$.

\begin{figure}[tbp]
     \begin{center}
     \includegraphics[width=3.4in,keepaspectratio]{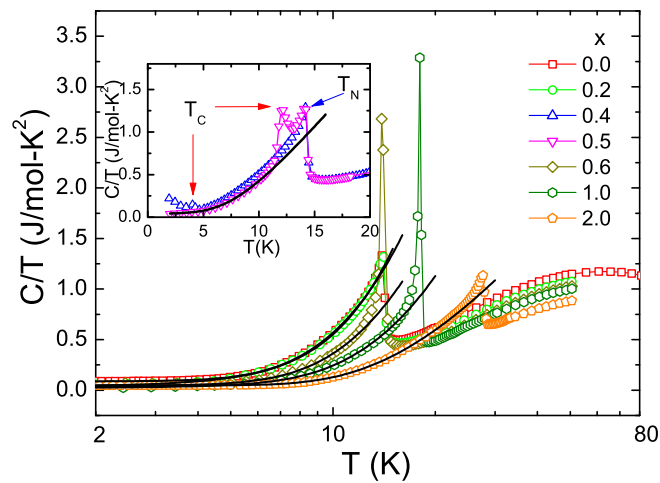}
     \end{center}
     \caption{Temperature dependence of the specific heat divided by temperature, $C(T)/T$, at zero field for CePd$_2$As$_{2-x}$P$_x$ ($0 \leq x \leq 2$). The inset plots the $C/T$ for $x = 0.4$ and $0.5$. The solid lines are fits to the data using Eq. 3.}
\end{figure}

\subsection{Phase diagram and discussion}

\begin{figure}[tbp]
     \begin{center}
     \includegraphics[width=3.5in,keepaspectratio]{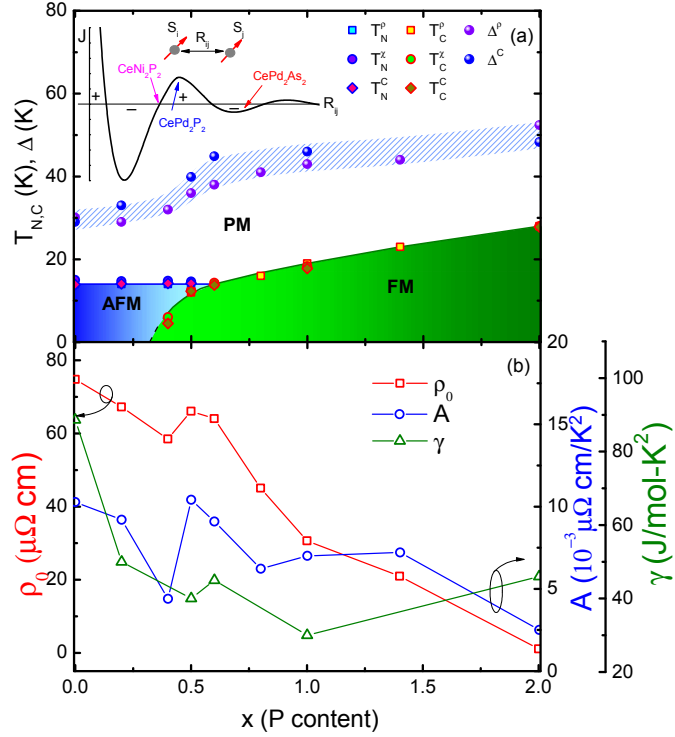}
     \end{center}
     \caption{(a) Magnetic phase diagram for CePd$_2$As$_{2-x}$P$_x$, constructed from  the magnetic susceptibility $\chi(T)$, specific heat $C(T)$ and electrical resistivity $\rho(T)$, together with the derived magnon gaps $\Delta(x)$. (b) The residual resistivity $\rho_0$ (left-axis), the resistivity $A$-coefficient and Sommerfeld coefficient $\gamma$ (right-axis) plotted as a function of the P-content $x$. The lines are guides to the eyes. Inset of (a) plots a variation of the exchange coupling constant $J$ as a function of distance $R_{ij}$. }
\end{figure}

Figure 8(a) presents the magnetic phase diagram of CePd$_2$As$_{2-x}$P$_x$ together with the spin gap $\Delta(x)$. The residual resistivity $\rho_0(x)$, the resistive coefficient $A(x)$ and the Sommerfeld coefficient $\gamma(x)$ are plotted as a function of the nominal P-content in figure 8(b). CePd$_2$As$_2$ undergoes an AFM transition at $T_N \approx $ 15 K, which is nearly unchanged with increasing the P-content up to $x \simeq 0.6$. On the other hand, a FM ground state develops at $T_C \approx $ 5 K for $x \simeq 0.4$; its $T_C$ increases with $x$ and approaches the N\'{e}el temperature at $x \simeq 0.6$. For $x \geq 0.6$, the compounds exhibit a typical FM behavior at low temperatures with the Curie temperature increasing with $x$ and reaching $T_C \approx$ 28 K in CePd$_2$P$_2$.

The spin gap $\Delta(x)$, derived from the specific heat and the electrical resistivity data, gives highly consistent values and shows a step-like increase around $x = 0.5$, indicating a sharp change of the spin gap between the AFM and FM states. Furthermore, the residual resistivity $\rho_0(x)$, the resistivity coefficients $A(x)$ and  Sommerfeld coefficient $\gamma(x)$ decrease with increasing the P-content, showing a weak anomaly around $x = 0.5$ [figure 8(b)]. For CePd$_2$As$_2$, the $\rho_0$ and $A$ coefficient are 74.78 $\mu\Omega$cm and 1.03$\times10^{-2}\mu\Omega$cm/K$^2$, respectively. In contrast, much smaller values are obtained for CePd$_2$P$_2$ ($\rho_0$ = 1.1 $\mu\Omega$cm and $A$ = 2.5$\times 10^{-3}\mu\Omega$cm/K$^2$). Our results demonstrate that the metallicity is enhanced with increasing $x$ in CePd$_2$As$_{2-x}$P$_x$.

The magnetic phase diagram of CePd$_2$As$_{2-x}$P$_x$ resembles that of EuFe$_2$(As,P)$_2$ from the aspect of $4f$-magnetism \cite{Cao2010}. In EuFe$_2$As$_2$, the $3d$- and $4f$-electrons undergo a magnetic/structural transition around 190 K and a Eu-AFM transition at 19 K, respectively \cite{Cao2010}. Similar to CePd$_2$As$_2$, a metamagnetic transition was also observed in EuFe$_2$As$_2$ at low temperatures \cite{Jiang2008}. Substitution of As with P in EuFe$_2$(As,P)$_2$ leads to the suppression of the Fe-AFM order and the emergence of superconductivity. On the other hand, the Eu-$4f$ electrons evolve from an AFM ground state to a FM state with increasing P-content, and its Curie temperature reaches $T_C \approx 29$ K in EuFe$_2$P$_2$ \cite{Cao2010}. The neutron scattering experiments suggest that the Eu-moments in the FM state are tilted from the $c$-axis by 17 degrees \cite{ryan2011}. Our measurements reveal that Ce-$4f$ electrons undergo an AFM transition at $T_N \approx$ 15 K in CePd$_2$As$_2$, which evolves into a FM state by applying either chemical pressure (As/P substitution) or magnetic field. The moderate Sommerfeld coefficient ($\gamma$ $\approx$ 88 mJ/mol-K$^2$) of CePd$_2$As$_2$ indicates a sizable hybridization between the $f$ electrons and conduction electrons. The evolution of the magnetic order as a function of As/P substitution in these compounds is attributed to the modulation of the Ruderman-Kittel-Kasuya-Yosida (RKKY) interaction tuned by the lattice distance. The RKKY interaction between two localized electrons of $S_i$ and $S_j$ can be expressed as: $H_{ij} \propto -J(R_{ij}) S_i \cdot S_j$, where $J(R_{ij})$ is the exchange integral between the i$th$ and the j$th$ spin separated by a distance of $R_{ij}$ [see inset of figure 8(a)] \cite{RKKY}. The sign of $J(R_{ij})$ oscillates from positive to negative as the separation $R_{ij}$ varies, leading to a change of magnetic order as a function of pressure. As schematically shown in the inset of Figure 8(a), the As/P substitution may tune the $J(R_{ij})$ from a negative value to a positive one, yielding a rich magnetic ground state in CePd$_2$As$_{2-x}$P$_x$. Further application of pressure to CePd$_2$P$_2$ may suppress its FM order to a QCP. Indeed, our preliminary data of Ce(Pd,Ni)$_2$P$_2$ reveal that the Pd/Ni substitution can suppress the FM transition in CePd$_2$P$_2$ and induce a HF state on the Ni-rich side. These observations imply that CePd$_2$P$_2$ is a potential candidate compound for the study of FM QCP. Meanwhile, it is also desirable to tune CePd$_2$As$_2$ or CePd$_2$P$_2$ by applying physical pressure, which would be an alternative and clean approach to study the possible superconductivity and magnetic quantum criticality in this system. Detailed measurements are still in progress.

\section{Conclusion}

In summary, we have successfully synthesized the polycrystalline compounds CePd$_2$As$_{2-x}$P$_x$ ($0 \leq x \leq 2$) and studied their physical properties by measuring the electrical resistivity, magnetic susceptibility and specific heat, from which a temperature-doping phase diagram is obtained. CePd$_2$As$_2$ undergoes an AFM-type magnetic transition at $T_N \approx$ 15 K. Either the As/P substitution or magnetic field tunes the ground state from an AFM order in CePd$_2$As$_2$ to a FM state. A FM ground state with a Curie temperature of $T_C$ = 28 K is observed in CePd$_2$P$_2$. Around a critical value of $x \simeq 0.5$ where the magnetic state switches from an AFM state to a FM one, the spin gap shows a step-like increase. Further investigations are needed to determine the  magnetic structure of CePd$_2$As$_{2-x}$P$_x$ and to study the quantum critical behavior in these compounds by applying pressure and/or magnetic field.

\ack{We would like to thank F. C. Zhang, M. B. Salamon and R. E. Baumbach for useful discussions. This work is partially supported by the National Basic Research Program of China (Nos.2009CB929104 and 2011CBA00103),the National Science Foundation of China (Nos.10934005, 11174245), Zhejiang Provincial Natural Science Foundation of China and the Fundamental Research Funds for the Central Universities.}

\end{document}